\documentclass[aps,prl,twocolumn,groupedaddress]{revtex4}
\usepackage{latexsym}
\newcommand{\ddt}{\ensuremath{\frac{\dr{}}{\dr{t}}}}
\newcommand{\ket}[1]{\ensuremath{\left|  #1 \right\rangle}}
\newcommand{\op}[1]{\ensuremath{\widehat{\textsf{\ensuremath{#1}}}}}
\newcommand{\opad}[1]{\ensuremath{\op{#1}^{\dagger}}}
\newcommand{\be}{\begin{equation}}
\newcommand{\ee}{\end{equation}}
\newcommand{\denop}{\ensuremath{\rho}}
\newcommand{\comm}[2]{\ensuremath{\left[ #1 , #2 \right]}} 
\newcommand{\dr}[1]{\ensuremath{\mathrm{d} #1\,}}
\newcommand{\matel}[3]{\ensuremath{\bra{#1} #2 \ket{#3}}}
\newcommand{\bra}[1]{\ensuremath{\left\langle #1 \right|}}
\newcommand{\vctr}[1]{\ensuremath{\mathbf{ #1 }}}

\begin{document}

\title{Non-locality and gauge freedom in Deutsch and Hayden's
formulation of quantum mechanics}
\author{David Wallace}
\email{david.wallace@philosophy.ox.ac.uk}
\affiliation{Magdalen College, University of Oxford}
\author{Chris Timpson}
\email{c.g.timpson@leeds.ac.uk}
\affiliation{School of History and Philosophy of Science, Department of Philosophy, University of Leeds}

\date{\today}

\begin{abstract}
Deutsch and Hayden have proposed an alternative formulation of quantum
mechanics which is completely local. We argue that their proposal must
be understood as having a form of `gauge freedom' according to which
mathematically distinct states are physically equivalent. Once this
gauge freedom is taken into account, their formulation is no longer
local.
\end{abstract}
\maketitle

\section{Introduction}

Unitary quantum mechanics (that is, quantum mechanics without collapse
of the wave function) has local interactions: the quantum state of a
system (e.g.\,a qubit, or a spacetime region in quantum field theory) is
affected only by influences which propagate via the quantum states of
its immediate past light cone.\footnote{In QFT, this is a consequence of
the requirement that spacelike separated observables must commute.}

As conventionally presented, though, QM does not have local
\emph{states}: if $S_1$ and $S_2$ are systems with quantum states $\denop_1$ and
$\denop_2$, then because of entanglement the state of the composite 
system $S_1 \times S_2$ is not necessarily $\denop_1 \otimes \denop_2$.

Deutsch and Hayden\cite{deutschhayden} argue that this `state nonlocality' 
is an artifact of the normal way in which we represent quantum states,
and that it disappears in an alternative formalism which they propose.
Their formalism is derived from the Heisenberg picture of quantum
mechanics, in which the unitary time evolution is applied to the
observables rather than to the state vector. In the normal understanding of
that formalism, though, the state vector is still taken to express the
physical state of the system (via its role in calculating expectation values) and the
algebra of observable quantities is regarded as mathematical `superstructure',
used to help us to calculate those observables.

Deutsch and Hayden reverse this `normal understanding'. They regard the
state vector \ket{0} as fixed, once and for all and independent of the physical
state of the system, and they regard the state of a quantum system as
literally given by the associated observables (so that 
the state of a qubit, for instance, is given by the triple of
Heisenberg picture operators $S_x, S_y, S_z$ pertaining to the spin
observables of that qubit). The dynamics of this theory are given by 
\be \label{truedyn}\ddt\op{X}_i= \frac{-i}{\hbar}\comm{\op{H}(\op{X}_1, \ldots
\op{X}_n)}{\op{X}_i}\ee
(where $\op{X}_1, \ldots \op{X}_n$ are the observables of the theory). 
It is easy to see that the theory is local in both the interaction and the
state senses, apparently vindicating Deutsch and Hayden's claims.

\section{Quantum gauge transformations}

Suppose $\op{V}(t)$ is a function from times to unitary operators, and
suppose that for each $t$, $\op{V}(t)\ket{0}=\exp(-i \theta) \ket{0}$
(for arbitrary phase factor $\theta$). Then if the state is represented,
according to Deutsch and Hayden, by observables $\op{X}_1, \ldots
\op{X}_n$, suppose that we make the transformation
\be \op{X}_i(t) \longrightarrow
\op{X}_i'(t)=\opad{U}(t)\op{X}_i(t)\op{U}(t).\ee
If $\op{V}(t)$ is not a constant then this changes the dynamics to
\be \label{gaugedyn}\ddt\op{X}'_i = \frac{-i}{\hbar}
\comm{\op{H}(\op{X}'_1, \ldots\op{X}'_n)}{\op{X}'_i} + \frac{-i}{\hbar}
\comm{\opad{V}(t)\ddt\op{V}(t)}{\op{X}'_i}.\ee
It does not, however, change anything observable, since everything
observable is given by the expectation values of observables with
respect to \ket{0}, and clearly
\be \matel{0}{\op{X}'_i}{0}=\matel{0}{\op{X}_i}{0}.\ee

To understand the significance of these `quantum gauge transformations',
it is useful to consider an analogous example: electromagnetism in the
context of the Aharonov-Bohm effect \cite{aharonovbohm}. Recall: the
electromagnetic potential \vctr{A} couples to electron wavefunctions
via the rule
\be \op{P}\longrightarrow \op{P}+e\vctr{A}.\ee
If an electron beam is split, passed on either side of a solenoid, and
recombined, there will be interference between the beams, and as the field in the
solenoid is varied the interference fringes will shift by an amount
proportional to the line integral of \vctr{A} around the electron's
path. This occurs despite the fact that the magnetic field outside the
solenoid is zero or nearly so.

The A-B effect
makes clear that the electromagnetic potential \vctr{A}, and not just
the fields \vctr{E} and \vctr{B}, must be regarded as physically
significant; however, all observable quantities (including the A-B
effect itself) are invariant under gauge transformations
\be \vctr{A}\longrightarrow \vctr{A}'=\vctr{A}+ \nabla f\ee
for arbitrary smooth functions $f$ (along with an associated
transformation of the wavefunction).

It is generally accepted that the correct response to this observation
is to regard gauge-equivalent \vctr{A}s as describing the same physical
situation, so as not to burden our theory with massive indeterminism
(caused by the possibility of arbitrary \emph{time-dependent} 
gauge transformations) and with an excess of unobservable properties
(caused by the fact that the observable data \emph{right now} only fixes
the state up to a gauge transformation).

However, this does come with a price: if we identify gauge-equivalent
vector potentials then our theory has non-local states in the sense
described above. For while the Aharonov-Bohm vector potential cannot be
gauge-transformed to zero everywhere, it can be in any region which does not
completely enclose the solenoid. Since a region which \emph{does}
enclose the solenoid can be decomposed into regions which do not, it
follows that whether the solenoid-enclosing region induces an A-B effect
is not determined by the properties of its parts.

The loop representation of \vctr{A} makes this state
non-locality manifest. We replace \vctr{A} with the \emph{loop phases}
\be C_\gamma = \int_\gamma \vctr{A} \cdot \dr{x}\ee
where $\gamma$ is any closed loop. \vctr{A} is fixed up to gauge
transformations by the $C_\gamma$, and $\vctr{B}_i$ is given at a point \vctr{x} by
the loop phase for an infinitesimal loop in a plane perpendicular to
$\vctr{e}_i$. A loop which encloses the solenoid cannot be expressed as
the sum of loops which do not enclose the solenoid, so the loop
representation has nonlocal states.

\section{Lessons for quantum mechanics}

The same arguments which lead us to identify gauge-equivalent vector
potentials should lead us to identify gauge-equivalent quantum states.
Specifically:
\begin{enumerate}
\item The possibility of time-dependent quantum gauge transformations
makes it undetermined which dynamical equations give the true dynamics
for the quantum state: is it (\ref{truedyn}) or some (\ref{gaugedyn})?
(\ref{truedyn}) is somewhat simpler, but it is unclear whether this is
sufficient: after all, in electromagnetism
\be \Box A_\mu =0\ee
is a somewhat simpler choice of dynamics than those given by many other
gauges, but this does not lead us to regard it as the `true' dynamics.
\item Even time-independent gauge transformations make the state grossly
underdetermined by observable data. Provided that $\op{V}\ket{0}=\exp(-i
\theta) \ket{0}$, nothing whatever --- no observable data, no
theoretical considerations --- can tell us that the physical state is
given by $\op{X}_1, \ldots \op{X}_n$ rather than
$\opad{V}\op{X}_1\op{V}, \ldots \opad{V}\op{X}_n\op{V}.$
\end{enumerate}
(There is also a more `philosophical' concern: in a physical theory we
would normally prefer that what is `observable' (\mbox{i.\,e.\,}, the expectation
values derived from \ket{0}) would emerge from a physical analysis of measurement,
rather than by \emph{fiat}.)

This suggests that we should identify Deutsch-Hayden states which differ
only by a gauge transformation. But if we do so, we return to the usual
representation of quantum states! For two Deutsch-Hayden states are
gauge-equivalent if and only if they have the same expectation values 
--- and of course the expectation values of all possible measurements on
a given quantum system are encoded in that system's density operator. So
if we do identify gauge-equivalent states, we are again left with a
theory whose states are non-local.

\section{Conclusion}

Deutsch and Hayden's proposal secures locality of states only at the
cost of a gauge freedom closely analogous to the gauge freedom of
electromagnetism. However, in quantum mechanics as in electromagnetism, to avoid
problems of indeterminism and state underdetermination it is necessary
to identify gauge-equivalent states. In quantum mechanics as in
electromagnetism, if we do make this identification then it leads to
nonlocality of states.

Deutsch and Hayden argue \cite[p.\,1772]{deutschhayden} that if a theory is
local according to any formulation, then it is local period. But their
version of quantum mechanics is only a new formulation if we do indeed
identify gauge-equivalent states. If not, it is not a `new formulation': it is a new
\emph{theory} --- with novel properties such as associating many distinct states
to the same in-principle-observable data --- 
albeit one which has the same observational consequences as the old
theory. (Deutsch has himself insisted on this distinction in his more
foundational work, for instance in discussing the de Broglie-Bohm
interpretation \cite{deutschlockwood}). It is a new theory which is
genuinely local, but which pays an unacceptably high price for that
locality.

We conclude that Deutsch and Hayden's proposal is best understood as a
gauge theory whose gauge-independent physical properties are given by
the normal quantum formalism. As such, although it may well give
important insights into quantum-information issues such as information
flow(for a detailed analysis of this point see \cite{timpson}), it does not achieve the goal of showing that quantum mechanics is
completely local. Rather, quantum mechanics has only local interactions, but has
nonlocal states.

\end{document}